\def\year{2021}\relax
\documentclass[letterpaper]{article} 
\usepackage{aaai21}  
\usepackage{times}  
\usepackage{helvet} 
\usepackage{courier}  
\usepackage[hyphens]{url}  
\usepackage{graphicx} 
\def\UrlFont{\rm}  
\usepackage{natbib}  
\usepackage{caption} 
\title{Embeddings-Based Clustering for Target Specific Stances:\\The Case of a Polarized Turkey }
\author {
    Ammar Rashed,\textsuperscript{\rm 1}
    Mucahid Kutlu,\textsuperscript{\rm 2}
    Kareem Darwish,\textsuperscript{\rm 3} 
    Tamer Elsayed,\textsuperscript{\rm 4} {\normalfont and}
    Cans{\i}n Bayrak\textsuperscript{\rm 2} \\
}
\affiliations {
    \textsuperscript{\rm 1} \"{O}zye\u{g}in University \\
    \textsuperscript{\rm 2} TOBB University of Economics and Technology
    \\
    \textsuperscript{\rm 3} QCRI, HBKU \\
    \textsuperscript{\rm 4} Qatar University \\
    ammar.rasid@ozu.edu.tr, m.kutlu@etu.edu.tr, kdarwish@hbku.edu.qa, telsayed@qu.edu.qa, c.bayrak@etu.edu.tr \\
}

\usepackage{xspace}
\usepackage{tabularx}
\usepackage{booktabs}
\usepackage{array,multirow}
\usepackage{subfigure}
\usepackage{stackengine}
\renewcommand{\prec}{\ensuremath{\mathcal{P}}\xspace}
\newcommand{\recall}{\ensuremath{\mathcal{R}}\xspace}
\newcommand{\fscore}{\ensuremath{\mathcal{F}}\xspace}

\begin{document}

\maketitle

\begin{abstract}
On June 24, 2018, Turkey conducted a highly consequential election in which the Turkish people elected their president and parliament in the first election under a new presidential system. During the election period, the Turkish people extensively shared their political opinions on Twitter. One aspect of polarization among the electorate was support for or opposition to the reelection of Recep Tayyip Erdo\u{g}an. In this paper, we present an unsupervised method for target-specific stance detection in a polarized setting, specifically Turkish politics, achieving 90\% precision in identifying user stances, 
while maintaining more than 80\% recall. The method involves representing users in an embedding space using Google's Convolutional Neural Network (CNN) based multilingual universal sentence encoder.
The representations are then projected onto a lower dimensional space in a manner that reflects similarities and are consequently clustered. We show the effectiveness of our method in properly clustering users of divergent groups across multiple targets that include political figures, different groups, and parties. We perform our analysis on a large dataset of 108M Turkish election-related tweets along with the timeline tweets of 168k Turkish users, who authored 213M tweets. Given the resultant user stances, we are able to observe correlations between topics and compute topic polarization.
\end{abstract}

\section{Introduction}
On June 24, 2018, Turkey conducted early elections for the presidency and the parliament that would bring into force the constitutional changes that were approved a year earlier.
The constitutional changes would transform Turkey from a parliamentary system to a presidential system. With the office of the president enjoying significantly-increased powers, these elections were considered highly consequential for Turkey. Nascent coalitions were formed in the lead up to the elections with several presidential candidates representing different Turkish political blocks such as conservatives, secularists, nationalists, and Kurds. Given the front runner status of the incumbent candidate, Recep Tayyip Erdo\u{g}an, and the newly-formed alliance between his party (AKParti) and the nationalist party (MHP), in this work, we explore the polarization between Twitter users who supported Erdo\u{g}an and those who favored other candidates using a novel fine-grained unsupervised stance detection method, which relies on mapping users into an embeddings space, projecting them to a lower dimensional space, and then clustering them. This enables us to further characterize polarization across different political and apolitical issues of interests to such users. 

For our analysis, we collected 108M election-related tweets between April 29 and June 23, 2018. Then, using semi-automatic labeling (based on self-declarations in users' profiles followed by a label propagation method), we labeled about 652.7K Twitter users, of which 279.2K are considered pro-Erdo\u{g}an and 373.5K are considered anti-Erdo\u{g}an. We estimate that tagging accuracy is above 95\%. Of those users, we crawled the timelines of 82K and 86K random users from pro- and anti-Erdo\u{g}an groups respectively to obtain tweets that were posted before and after the election. Timeline crawling yielded 213M tweets.

Previous work has suggested that word embeddings can capture human biases \citep{caliskan2017semantics} and hence determine shifting and/or divergent attitudes \citep{garg2018word,giatsoglou2017sentiment}.  

We build on previous work on embeddings to capture fine-grained divergences between polarized groups across political 
issues to ascertain if these divergences are persistent or transient. Specifically, we employ subword-level Convolution Neural Network (CNN) embeddings to map users, based on the text of their tweets about a specific topic, into an $n$-dimensional embeddings space.
As Turkish is an agglutinative language in which words can be very long due to many suffixes (e.g. predicative pronouns, plural markers, cases, etc.\footnote{\url{https://en.wiktionary.org/wiki/Appendix:Turkish_suffixes}}), using subword-based models can help overcome the morphological complexity of Turkish. Next, we project users into a lower dimensional space in a manner that brings similar users closer together and dissimilar users further apart, where similarity is computed based on their embeddings vectors. Lastly, the projected users are clustered. One advantage of this method is that it does not rely on any Twitter specific features. The resultant clusters can be contrasted with the positions of the users in the lead up to the elections, where users either supported Erdo\u{g}an or one of his rivals. We can quantify polarization on a topic, which allows us to determine if divergences on specific topics are transient or more systemic, and whether they are pragmatic or ideological. We examine topics relating to prominent politicians and popular political issues.
Our contributions in this work are four-fold:
\begin{itemize}
    \item We collected a large collection of tweets related to the Turkish election containing more than 108M tweets, and used semi-supervised methods to accurately tag more than 652k users given limited manual-tagging. We then collected more than 213M tweets from timelines of users from pro- and anti-Erdo\u{g}an groups.
    \item Given embeddings representations of user tweets on specific topics, we project and cluster users to determine if they are polarized on those topics. This allows us to measure cross-topic mutual information to determine the alignment of polarization across multiple topics. 
    \item We couple our projection with hierarchical clustering to further identify sub-groups, enabling us to gauge if stances align on specific sub-topic but diverge otherwise.
    \item Using embeddings with projection and clustering, we provide a comprehensive framework to analyze fine-grained polarization between groups over various topics, and we show the efficacy of this approach in performing unsupervised stance detection in general
\end{itemize}

\section{Related Work}
\paragraph{Stance Detection} 
Stance detection can be performed
using supervised classification and using a variety of features such as text-level features (e.g., words or hashtags), user-interaction features (e.g., user mentions and retweets), and profile-level features (e.g., name and location) \citep{borge2015content,magdy2016isisisnotislam,magdy2016failedrevolutions}. 
The use of retweets seems to yield competitive results \citep{magdy2016isisisnotislam,wong2013quantifying,wong2016quantifying}. 
Label propagation is also an effective semi-supervised method that propagates labels in a network based on follow or retweet relationships \citep{borge2015content,weber2013secular} or the sharing of identical tweets \citep{darwish2018scotus,kutlu2018devam,magdy2016isisisnotislam}. In this paper, we use an iterative label propagation method based on retweeted tweets. Other methods for user stance detection include: collective classification \citep{duan2012graph}, where users in a network are jointly labeled, and projecting users into a lower dimensional user space prior to classification \citep{darwish2017improved}.  More recent work projects user onto a two dimensional space and then uses clustering to perform unsupervised stance detection \citep{darwish2019unsupervisedStance}.  In their work, the best setup used UMAP for projection, mean shift for clustering, and the retweeted accounts as user features.  We employ a similar approach with two main differences, namely: we represent the content of user tweets on specific topics using CNN embeddings vectors, availing the need for Twitter specific features such as retweets; and we employ HDBSCAN for clustering instead of mean shift as it seems to work better with our projected vectors and provides hierarchical clusters.  \cite{becatti2019extracting} described an unsupervised method for discovering ``alliances'' in  networks of verified and unverified accounts via the use of bipartite graphs.  They applied their method on the 2018 Italian elections.

\paragraph{Polarization on Twitter} 
\label{sec:quantifyingPolarization}
Social media is a fertile ground for polarization, due to two social phenomena, namely: homophily, which is the tendency of similar users to congregate together, and biased-assimilation, where individuals readily accept evidence confirming their group's view, but are rather critical when provided with disconfirming evidence. Both phenomena are amplified on social network platforms \citep{bias_assim,homophily,homophily_twitter}.
 
Further, online social networks facilitate discovery and communication between like-minded users and hence the creation of large homophilous communities \citep{homophily_twitter}. Biased-assimilation has been shown to play a crucial role in the dynamics of polarization, as it makes community members more entrenched in their views, particularly on controversial topics \citep{bias_assim}. 
The dynamics of \textit{intra-community} and \textit{inter-community} interactions provide predictive information about potential conflicts \citep{weber2013secular}. \cite{kumar2018community} introduced a method that employs graph embeddings, where the graph captures user interactions on Reddit\footnote{https://www.reddit.com/}, in conjunction with user, community, and text features to predict potential conflict and subsequent community mobilization. Unlike Twitter, the pseudo-anonymous nature of Reddit users may affect the types of interactions between communities. 
Other work has focused on quantifying polarization \citep{darwish2019quantifying,garimella2018quantifying,guerra2013measure,morales2015measuring}.  Several methods were used including random graph walks, network betweenness, distances in embedding spaces for different groups \citep{garimella2018quantifying}, inter-group and intra-group distances \citep{morales2015measuring}, and popularity of boundary nodes between communities \citep{guerra2013measure}. Given polarized communities, several studies looked at identifying distinguishing features, such as hashtags, between such communities. One method uses the so-called valence score that measures the relative probability of a feature appearing in one community compared to another \citep{conover2011political,darwish2018scotus,weber2013secular}. 

Word embeddings were shown to capture implicit human biases \citep{caliskan2017semantics}. Thus, several studies utilized word embeddings, trained on the content generated by different communities, to contrast them through the usage of similar words and words associated with different concepts \citep{garg2018word,giatsoglou2017sentiment}.   \cite{garg2018word} trained temporal word embeddings that span 100 years to measure shifts in racial and gender attitudes. They also correlated key concepts with positive and negative adjectives over time.  
\cite{giatsoglou2017sentiment} trained a polarity classification model using word embeddings with a seed lexicon of polarity-labeled words.
\cite{yang2017quantifying} used word2vec embeddings and clustering metrics to quantify polarization. 
\cite{an2019political} utilized semantic differences to analyze interaction patterns in and between homogeneous groups. They quantified semantic difference as the cosine distance between words' vector representation between different but aligned word embeddings trained on each groups' vocabulary.

We extend prior work on using embedding representations of tweets in multiple ways, namely: we use pre-trained CNN embeddings going beyond word boundaries to allow us to project groups with divergent groups into a unified embeddings space; we specifically use subword-segment embeddings to overcome the morphological complexities of Turkish (such can be helpful to other morphologically rich languages); and we show that we can construct embeddings vector representations of users on specific topics and subsequently automatically cluster users with high accuracy. Our model does not need manual labeling  
to determine users' polarization on a specific topic.

\section{Background: Turkish Elections}
Through a referendum on April 16, 2017, Turkey made significant changes in its constitution, effectively changing the government from a parliamentary system to a presidential system, giving more power to the president. 
The Turkish president Erdo\u{g}an announced the first election under the new constitution that was held on June 24, 2018 to elect both the president and parliament simultaneously.
Voter participation was 86.24\% with eight political parties participating in the parliamentary elections.
For the first time in the history of Turkish elections, parties were also allowed to make alignments in parliamentary elections, such as the ``Public Alignment'', which included Justice and Development Party (AKParti) and Nationalist Movement Party (MHP), and the ``Nation Alignment'', which included the Republican People's party (CHP), the Good Party (IYI) and Felicity Party (SP). Such alignments brought parties with different ideological backgrounds together. In the presidential elections, there were 5 candidates from these major parties, namely Recep Tayyip Erdo\u{g}an (AKParti), Muharrem Ince (CHP), Selahattin Demirta\c{s} (Peoples' Democratic Party (HDP)), Meral Ak\c{s}ener (IYI), and Temel Karamollao\u{g}lu (SP). 
MHP and Huda-Par, a minor Kurdish Islamist Party, announced their support for Erdo\u{g}an in the presidential election.

The incumbent and front runner status of Erdo\u{g}an caused voters to cast the elections as referendum on continuing his presidency or not. Hence the hashtag \#devam (``continue'') became popular among his supporters, while his opponents, regardless of political affiliation, used the hashtag \#tamam (``enough''). In our data also, we have seen that many users used these hashtags, not just in their tweets, but also in their screen names. Therefore, Turkish voters can be roughly divided into two groups: pro and anti-Erdo\u{g}an voters.  We have also observed this political binarization in Turkish politics in the last elections held on 31 March, 2019, with previous alignments holding. 
Though HDP, commonly associated with Turkey's Kurdish minority, was not a part of the Nation Alignment and did not field candidates for many cities, HDP announced support for the most favorable candidate running against the Public Alignment's candidate in such cities\footnote{\url{www.bbc.com/turkce/haberler-turkiye-48213123}}.    

\section{Dataset}
We constructed two different datasets for our study. First, we collected election-related tweets, denoted as the \textbf{Election Dataset} (\textbf{ED}). Next, we labeled the users as pro-Erdo\u{g}an (Pro) or anti-Erdo\u{g}an (Anti) using a two-step label-propagation approach. 
Subsequently, we crawled all tweets of randomly selected labeled users to construct our \textbf{Timeline Dataset} (\textbf{TD}).

\paragraph{\textbf{Election Data Crawling:}}\label{section_election_time_crawling}
We collected tweets related to Turkey and the election starting on April 29, 2018 until June 23, 2018, which is the day before the election. We tracked keywords related to the election including political party names, candidate names, popular hashtags during this process (e.g., \#tamam and \#devam), famous political figures (e.g., Abdullah G\"{u}l, the former president of Turkey), and terms that may impact people’s vote (e.g., economy and terrorism). We wrote keywords in Turkish, which contains some additional letters that do not exist in the English alphabet (e.g., \c{c}, \u{g}, \c{s}). Next, we added versions of these keywords written strictly with English letters (e.g., ``Erdogan'' instead of ``Erdo\u{g}an''), allowing us to catch non-Turkish spellings. Overall, we collected 108M tweets. 

\paragraph{\textbf{Labeling:}}\label{sec_labeling} 
The labeling process was done in two steps:\\

\textbf{(1) Manual labeling:} First, we assigned labels to users in ED set who explicitly specified their party affiliation in their Twitter handles or screen names. We made one simplifying assumption, namely that supporters of a particular party would be supporting the candidate supported by their party. We extracted a list of users who use ``AKParti'', ``CHP'', ``HDP'', or ``IYI'' in their Twitter handles or screen names. We labeled the users who used ``AKParti'' as ``pro-Erdo\u{g}an'', while the rest as ``anti-Erdo\u{g}an''. Though ``MHP'' officially supported Erdo\u{g}an in the election, we suspected that MHP supporters might not be universally supporting Erdo\u{g}an due to various news articles and surveys.\footnote{\url{https://konda.com.tr/wp-content/uploads/2018/05/KONDA_SecmenKumeleri_MHP_Secmenleri_Mayis2018.pdf}} Consequently, we did not label MHP supporters as `pro-Erdo\u{g}an''. 
Further, we labeled users who had the hashtags \#devam or \#tamam in their profile description as supporting or opposing Erdo\u{g}an respectively. Lastly, users who had the hashtag \#RTE (i.e., initials of Recep Tayyip Erdo\u{g}an) in their profile description were labeled as pro-Erdo\u{g}an. While providing a political party name as a part of Twitter user profile is a strong indication of supporting the respective party, we \emph{manually} checked all extracted names to ensure the correctness of labels. For instance, we found that some users expressed that they are against a particular party in their user name instead of supporting it. Therefore, whenever we suspected that keywords we used for labeling were not indicative of their political view, we manually investigated the accounts and removed their labels if their political views were unclear. The total number of manually labeled users are listed in \textbf{Table~\ref{tab:proAntiErdogan}}. 

\noindent\textbf{(2) Label Propagation:} Next, we automatically labeled users based on the tweets that they retweeted \citep{darwish2017predicting,darwish2018scotus}. The intuition behind this method is that users retweeting the same tweets most likely share the same stances on the topics of the tweets. Given that many of the tweets in our collection were actually retweets, we labeled users who retweeted 10 or more tweets that were posted or retweeted exclusively by the pro- or anti- groups and no retweets from the other side as pro- or anti- respectively. We iteratively performed such label propagation 11 times, which is when label propagation stopped labeling new accounts. By the last iteration, we had labeled 652,729 users of which 279,181 were pro-Erdo\u{g}an and posted 28,050,613 tweets, and 373,548 were anti-Erdo\u{g}an and posted 31,762,639 tweets. 
This label propagation method imposes strict conditions to avoid leakage from one class to the other.  However in doing so, it generally labels users with potentially more entrenched stances.
To ensure labeling accuracy, we manually and independently labeled 100 users from each of the pro and anti groups, and found that label propagation matched manual labeling for 191 users; 1 label was clearly wrong, and we could not decide on the stances of the 8 remaining users due to insufficient political tweets. 

\begin{table}[tbp]
\begin{center}
\begin{tabularx}{\linewidth}{X|r|r}  

\textbf{Supporters} &  \textbf{Users}  & \textbf{Tweets}\\  \hline
pro-Erdo\u{g}an & 1,772 & 561,510\\ \hline
anti-Erdo\u{g}an w/o party affiliation & 2,115 & 516,166\\ \hline
pro-CHP & 29 & 171,201 \\ \hline
pro-IYI & 890 & 168,442\\ \hline
pro-HDP & 354 & 61,274\\ \hline
\multicolumn{1}{r|}{\textbf{Total}} & 5,960 & 1,478,593 \\
\hline
		\end{tabularx}
\end{center}
    \caption{Manually-labeled users}
    \label{tab:proAntiErdogan}
\end{table}

\paragraph{\textbf{Timeline Data Crawling:}}\label{sec_timeline_crawling} 
On Dec. 28, 2018, we started crawling the timelines of 86,116 and 81,963 pro and anti users respectively using Tweepy\footnote{\url{http://www.tweepy.org/}}, which is a Twitter API wrapper. Though we started with an equal number of users for both groups, some of the user accounts were either closed or suspended at the time of crawling.  Twitter typically allows the crawling of the last 3,200 tweets for a user. Depending on how active each user is, 3,200 tweets can cover days, months, or years. We also excluded all non-Turkish tweets, where we relied on the language tag Twitter provides. In all, we collected 98,700,529 and 115,047,039 tweets from pro and anti groups, respectively, with some of the tweets dating back to 2013.

\section{Data Pre-processing}

Due to the informal nature of Twitter, tweets commonly have grammatical and spelling errors.  Furthermore, 
Twitter users frequently use emojis, emoticons, hashtags, media links, and other non-alphabetic characters. 
Thus, we performed the following pre-processing steps for the tweets on the ED and TD sets:\\
-- Case folding, where we lower-cased letters.\\
-- Removal of all links and user mentions.\\
-- Removal of all non-letter characters and punctuation. \\
-- Replacement of all numbers to the word ``number''. \\
-- Noisy text normalization.\footnote{We used the normalization function of Zemberek-nlp library \url{https://github.com/ahmetaa/zemberek-nlp}}

\section{Embeddings-based Stance Detection Approach}
\subsection{Description of Our Approach}
As stated earlier, we aim to perform unsupervised fine-grained user-level stance detection based on the content of user tweets. Specifically, our proposed method represents users in an embeddings space using their topically-relevant tweets, projects user representations to a lower dimensional space, and then clusters users. 
\paragraph{Tweet and User Representations} Embeddings are able to capture syntactic and semantic knowledge about words and word sequences \citep{garg2018word}. In this work, we use Google's Multilingual Universal Sentence Encoder (MUSE) with pre-trained CNN embeddings \citep{yang2019multilingual} to represent tweets. The embeddings were trained on a large multilingual corpus containing text in 16 languages including Turkish. The text was tokenized using SentencePiece (a.k.a. BP), which produces sub-word tokens \citep{kudo2018sentencepiece}. MUSE takes a sentence as an input sequence, tokenizes it, and produces a 512 dimensional output vector. MUSE has been shown to produce competitive results for a variety of natural language processing and retrieval tasks \citep{yang2019multilingual}.\footnote{\url{https://tfhub.dev/google/universal-sentence-encoder-multilingual/3}}  
Given a target of interest, such as the name of a political party or the name of a politician, we filter user tweets to obtain all tweets that mention the target. 
Next, we pass all the filtered tweets to MUSE to obtain vector representations of each.  To represent individual users, we take the average of all the vectors of the filtered tweets that were posted by the user. 

\paragraph{User Projection} We then project each user vector onto a two-dimensional plane using Uniform Manifold Approximation and Projection (UMAP) algorithm \citep{mcinnes2018umap}.\footnote{We used the UMAP-learn library at \url{https://umap-learn.readthedocs.io/en/latest/}} UMAP attempts to project the data elements in a manner that reflects the similarity between them, such that more similar elements are placed closer together and less similar elements are placed further apart. UMAP is more computationally efficient than other projection techniques, such as the Force Directed graph drawing technique \citep{fruchterman1991graph} and t-distributed Stochastic Neighbor Embedding (t-SNE) \citep{maaten2008visualizing} and generally produces better projections~\citep{mcinnes2018umap}. 
\paragraph{Clustering} We then cluster the projected user vectors using hierarchical density based clustering (HDBSCAN)~\citep{mcinnes2017accelerated}, which finds clusters of varying densities.\footnote{We used the hdbscan library at \url{https://pypi.org/project/hdbscan/}} We elect to project users prior to clustering, because clustering is typically less effective and less efficient in high-dimensional spaces. Clustering in high dimensional spaces suffers from the curse of dimensionality~\citep{VerleysenCurseDim2003}, and the distances between points, whether similar or dissimilar, begin to converge, which adversely affects clustering~\citep{beyer1999when}. We also conducted side experiments, where we tried to cluster users without projection, and the clustering results were poor.

\subsection{Validating Our Approach}
Before applying our proposed method on the Turkish election data, we validated its efficacy by comparing to existing supervised and unsupervised methods on two datasets for which we have ground truth labels. In the following, we provide the details of our experiments.

\paragraph{Datasets}  
We used two datasets for evaluation.  The first comprises the 5,981 users who were manually labeled as pro- or anti-Erdo\u{g}an (\textbf{Table~\ref{tab:proAntiErdogan}}) along with their 53,185 tweets in the ED dataset.  The second is a dataset containing US Twitter users who were labeled as either pro- or anti- the American president Donald Trump \citep{darwish2019unsupervisedStance}. Based on tweets that were collected in the span of three days (Oct. 25-27, 2018), the dataset has $13,731$ users, with $7,421$ and $6,310$ labeled as pro- and anti-Trump respectively. The accounts were labeled based on stance-indicative hashtags in the profile descriptions of the users, where users with the hashtag \texttt{\#MAGA} (Make American Great Again) were labeled as pro-Trump, and those with any of the hashtags\texttt{\#resist}, \texttt{\#resistance}, \texttt{\#impeachTrump}, \texttt{\#theResistance}, or \texttt{\#neverTrump} were labeled as anti-Trump. In all, pro- and anti-Trump users posted 166,814 and 148,178 tweets respectively.

\paragraph{Baselines}
We compared our method to supervised and unsupervised setups.
For the supervised method, we represented each user using a vector of all unique accounts that the user retweeted. Then, we used a Support Vector Machine (SVM) classifier to perform classification.\footnote{We used the Scikit learn SVC implementation:~\url{scikit-learn.org/stable/modules/generated/sklearn.svm.SVC}} This is consistent with the best setup used by~\cite{magdy2016isisisnotislam} and~\cite{darwish2017predicting}.  We ran two setups of this supervised method. In the first (denoted as \textit{SVM}$_{200}$), we used the set of the 200 most vocal users (ones with the most tweets) for training and the remaining users for testing.  This is to emulate a typical situation where a set of users is manually annotated.  In the second (\textit{SVM}$_{xval}$), we used five-fold cross validation on all the tagged users.  Though the availability of hundreds or thousands of accounts is not very common in practice, this would likely give us an upper bound on the effectiveness of this method.  
 
For the unsupervised method (denoted as \textit{UnSup}), we used the method proposed by~\cite{darwish2019unsupervisedStance}, which represents each user using a vector composed of all unique accounts that the user retweeted, projects the users onto a two dimensional space using UMAP, and clusters using mean shift clustering. Since unsupervised methods essentially perform clustering, they could potentially produce an arbitrary number of clusters and not necessarily 2 clusters. Hence, we assigned a label to each cluster based on the majority class in that cluster. For example, if the majority of users in a cluster were anti-Erdo\u{g}an, then we assigned this label to every user in the cluster and incorrectly labeled users would lower the overall precision.

\begin{table}[tbp]
    \centering
    \begin{tabular}{l|l|r|r|r||r|r|r}
    \multicolumn{2}{l}{} & \multicolumn{3}{c||}{\textbf{Trump}} & \multicolumn{3}{c}{\textbf{ED}} \\
\multicolumn{2}{l|}{} & PRO & ANTI & AVG & PRO & ANTI & AVG \\\hline

& \prec & 0.84	&	0.97	&	0.91	&	0.49	&	1.00	&	0.74 \\\cline{2-8}
& \recall    &  0.98	&	0.78	&	0.88	&	1.00	&	0.58	&	0.79 \\\cline{2-8}
\multirow{-3}{*}{\rotatebox[origin=c]{90}{\small{\textit{SVM}$_{200}$}}}& \fscore  & 0.91	&	0.86	&	0.89	&	0.65	&	0.74	&	0.69 \\\hline \hline
& \prec & 0.96	&	0.91	&	0.94	&	0.99	&	0.90	&	0.95 \\\cline{2-8}
& \recall    &  0.93	&	0.96	&	0.95	&	0.74	&	1.00	&	0.87 \\\cline{2-8}
\multirow{-3}{*}{\rotatebox[origin=c]{90}{\small{\textit{SVM}$_{xval}$}}}& \fscore  & 0.94	&	0.93	&	0.94	&	0.85	&	0.95	&	0.90 \\\hline \hline
& \prec & 0.94	&	0.93	&	0.94	&	0.99	&	0.98	&	0.99 \\\cline{2-8}
& \recall & 0.72	&	0.77	&	0.75	&	0.79	&	0.73	&	0.76 \\\cline{2-8}
\multirow{-3}{*}{\rotatebox[origin=c]{90}{\small{\textit{UnSup}}}} & \fscore & 0.82	&	0.84	&	0.83	&	0.88	&	0.84	&	0.86 \\\cline{1-8}  \hline \hline

& \prec & 0.90	&	0.87	&	0.89	&	0.88	&	0.92	&	0.90 \\\cline{2-8}
& \recall& 0.82	&	0.85	&	0.84	&	0.72	&	0.85	&	0.79 \\\cline{2-8}
\multirow{-3}{*}{\rotatebox[origin=c]{90}{\small{\textit{Ours}}}} & \fscore  & 0.86	&	0.86	&	0.86	&	0.79	&	0.89	&	0.84 \\\cline{1-8}

    \end{tabular}
    \caption{Benchmark results for Trump and Erdo\u{g}an datasets.}
    \label{tab:svc_baseline_trump}
\end{table}
\paragraph{Results}
\textbf{Table~\ref{tab:svc_baseline_trump}} compares the results of our method with the baselines using precision (\prec), recall (\recall), and F1 measures (\fscore). The results show that using more training data for the SVM classifier (for \textit{SVM}$_{xval}$) improves results over using less data (\textit{SVM}$_{200}$). The \textit{UnSup} method generally yielded high precision that is often higher than using the supervised methods; however, it led to much lower recall. Our method, where users are represented using MUSE embedding vectors, was competitive with the baseline supervised methods with the distinct advantage of being completely unsupervised.  When compared to \textit{UnSup}, though our method led to lower precision, it led to higher recall, leading to a comparable overall average F1-score. Given the results, we can see that our method has several advantages, namely: it is unsupervised; it does not rely on any Twitter specific features such as retweets; and, as we will see shortly, it can lead to finer-grained clustering that is difficult to achieve with the baseline methods.

\begin{figure}[tbp]
    \centering
    \includegraphics[width=.95\linewidth]{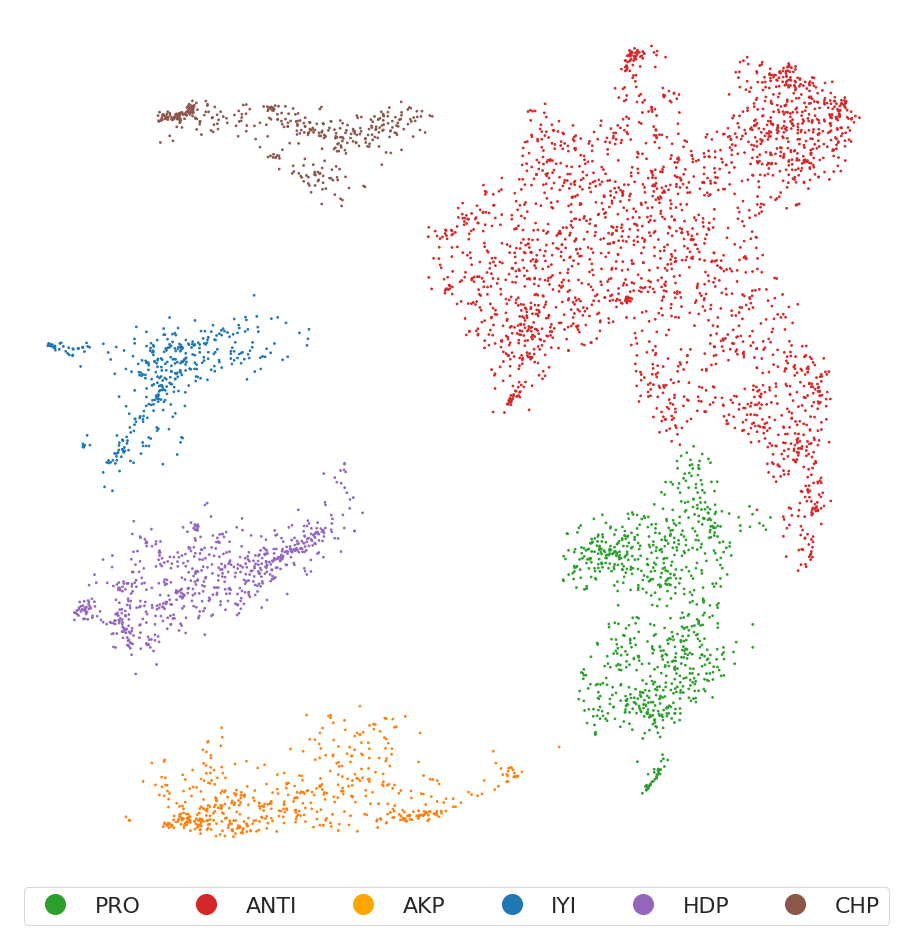}
    \caption{Party affiliation clusters on manually annotated users extracted by our method from ED dataset.}
    \label{fig:subclusters}
\end{figure}

It is noteworthy that although our method and the \textit{UnSup} method both produced 2 clusters for the Trump dataset, they in fact produced 6 and 3 clusters respectively for the ED dataset. Upon inspecting the three clusters produced by the \textit{UnSup} method, we found that each cluster was dominated by either pro-Erdo\u{g}an, anti-Erdo\u{g}an (without party affiliation), or pro-IYI party users.  However, using our method, the 6 clusters were dominated by either: pro-Erdo\u{g}an, anti-Erdo\u{g}an (without party affiliation), pro-CHP, pro-IYI, or pro-HDP users. We further manually inspected the two clusters that were dominated by pro-Erdo\u{g}an users, and we found that pro-AKP accounts were dominant in one cluster while the other cluster had mostly pro-Erdo\u{g}an users who did not explicitly express party affiliation. Therefore, in essence, our method was able to tweak apart all the constituent sub-groups in our dataset. \textbf{Figure~\ref{fig:subclusters}} shows the projection of the users using our method onto a two dimensional space, where we color-coded users according to their position/party affiliation. We also computed the precision, recall, and F1 of the 5 sub-groups for which we have gold labels (see \textbf{Table~\ref{tab:proAntiErdogan}}).  The results are shown in \textbf{Table~\ref{tab:ed_party}}.  As can be seen, the clusters generally have high precision (0.83 on average). We suspect that our method was able to more effectively cluster users at a finer-grained level compared to the \textit{UnSup} method, because the latter relies strictly on retweeted accounts and users may retweet a tweet if it agrees with their stance on a specific topic regardless of the political affiliation of the source. Conversely, our method uses the content of the tweets, and users with similar ideological stances may use similar language in their tweets.  We plan to investigate this further in future work.

\begin{table}[t]
    \centering
    \begin{tabular}{lrrrrrr}
\toprule
{} &   PRO &  ANTI &   HDP &   IYI &   CHP & Avg.  \\
\midrule
\prec &  0.88 &  0.74 &  0.68 &  0.94 &  0.90 & 0.83 \\\hline
\recall    &  0.72 &  0.83 &  0.77 &  0.69 &  0.43 & 0.69 \\\hline
\fscore  &  0.79 &  0.78 &  0.72 &  0.80 &  0.59 & 0.74 \\\hline
\bottomrule
\end{tabular}

    \caption{Aligning clusters with party affiliations with the manually labeled users on the ED dataset.}
    \label{tab:ed_party}
\end{table}

\section{Quantitative Analysis}
In this section, we utilized our method to quantitatively analyze
fine-grained topics. For our analysis, we picked 8 topics, which cover polarizing issues and personalities. We extracted the tweets of each topic from the TD dataset with indicative keywords. The topics with resultant tweets are described in \textbf{Table \ref{tab:topics}}. We ran our clustering method on all the users who talked about the topics.  We attempted to determine: 1) if we can infer the stance on one topic based on the stance on another topic; and what polarization quantification informs us about different topics.

{The overlap of a label, $l$, is defined as the fraction of users labeled \textit{and} predicted as $l$, over the entire set of users labeled \textit{or} predicted as $l$. In effect, this is Jaccard similarity.}

\begin{table}[tbp]
    \centering
    \small
\begin{tabularx}{\linewidth}{X|X|r|r}
Topic & Keywords & Users & Tweets \\ \hline
Arabs & Arap (Arab) & 39,918 & 858,237 \\ \hline
    CHP (main opposition party) & chp and k{\i}l{\i}\c{c}daro\u{g}lu (its party leader's surname) & 117,372	& 3,377,230 \\ \hline
    Erdo\u{g}an & erdo\u{g}an  & 131,389 & 5,203,924 \\ \hline
    HDP (Kurdish party) & hdp  and k\"{u}rt (Kurd) & 67,590	& 2,108,681 \\ \hline
    PKK (Kurdish militia) & pkk and ypg & 101,845 & 2,024,406 \\ \hline
    Syrians and refugees &  suriye (Syria) and m\"{u}lteci (refugee) & 112,459	& 1,688,988\\ \hline
    Trump & trump & 72,532 & 431,563 \\ \hline
    USA & amerika (America) and abd (USA) & 75,888 & 2,253,195\\ \hline
\end{tabularx}
\caption{Eight topics used for quantitative analysis.}
    \label{tab:topics}
\end{table}

\begin{figure}[tbp]
    \includegraphics[width=0.85\linewidth]{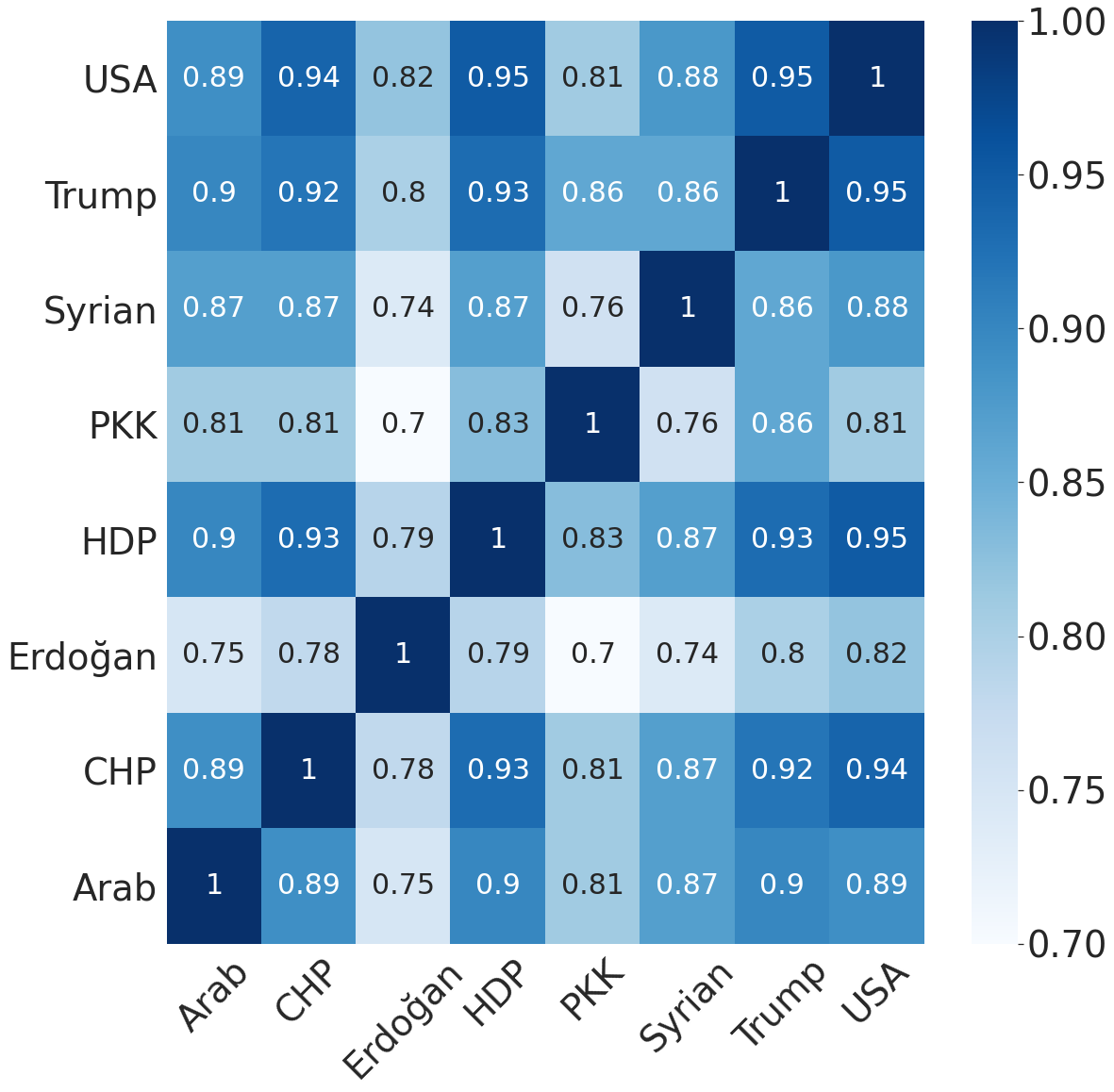}
    \caption{Adjusted mutual information between the clusters of different topics.}
    \label{fig:nmi}
\end{figure}

\begin{figure*}[pt]
    \centering 
    \subfigure[CHP]{\includegraphics[width=0.35\linewidth]{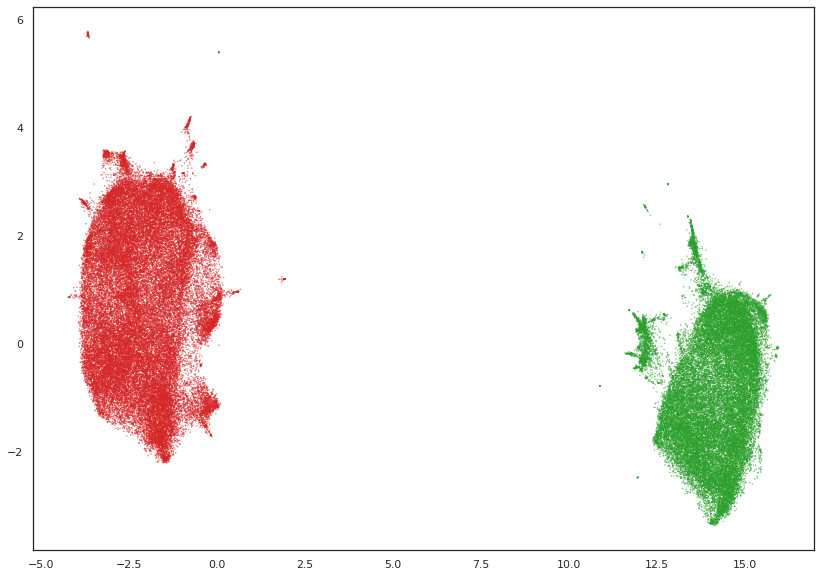}}
    \subfigure[Erdo\u{g}an]{\includegraphics[width=0.35\linewidth]{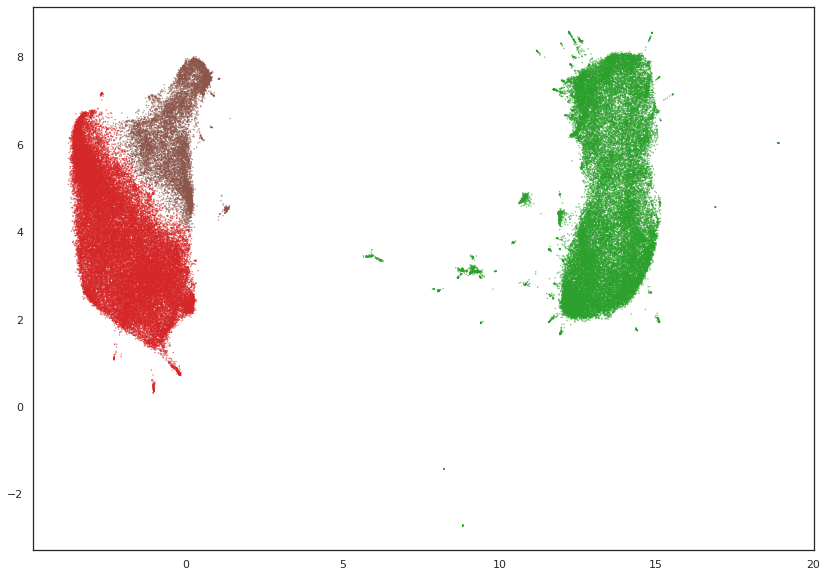}}
    
    \subfigure[Syria]{\includegraphics[width=0.35\linewidth]{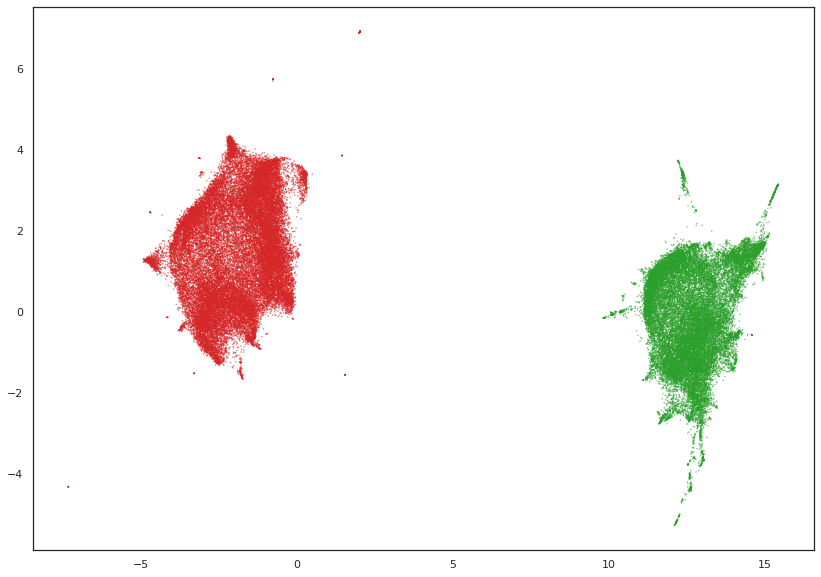}}
    \subfigure[Arab]{\includegraphics[width=0.35\linewidth]{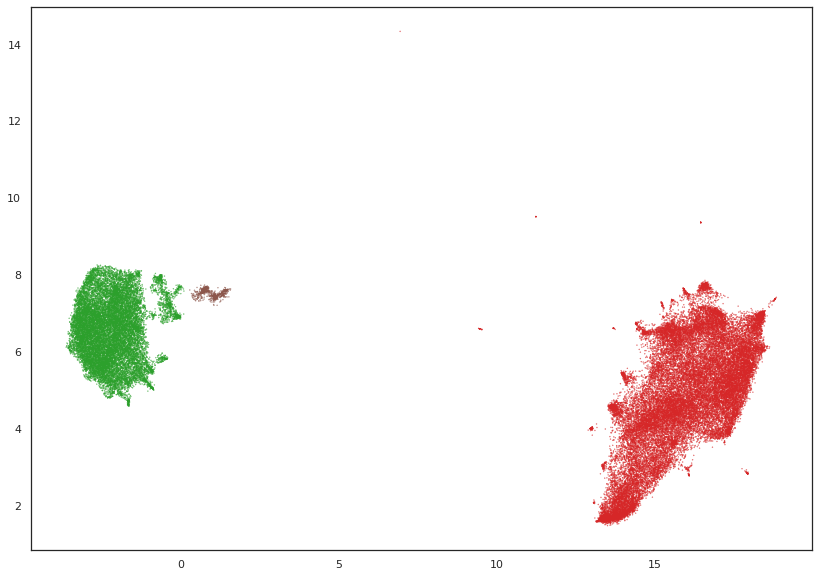}}
    
    \subfigure[US]{\includegraphics[width=0.35\linewidth]{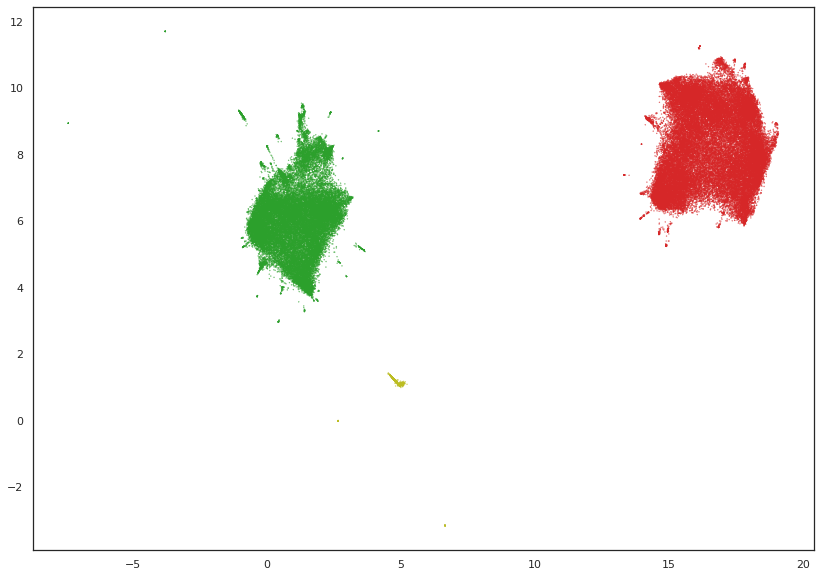}}
    \subfigure[Trump]{\includegraphics[width=0.35\linewidth]{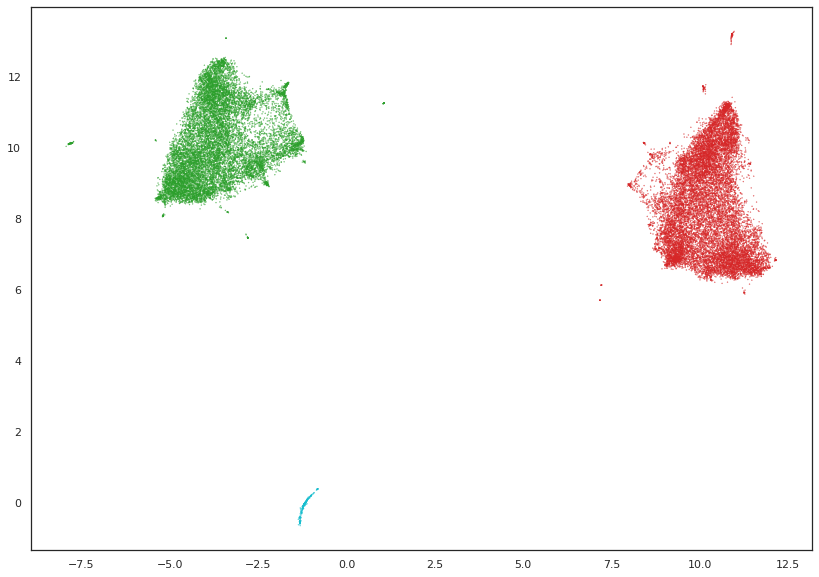}}
    
    \subfigure[PKK]{\includegraphics[width=0.35\linewidth]{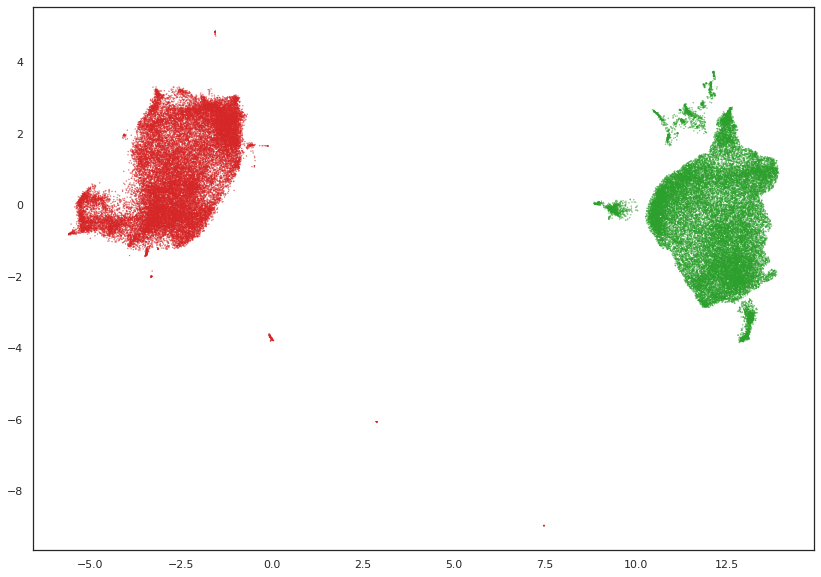}}
    \subfigure[HDP]{\includegraphics[width=0.35\linewidth]{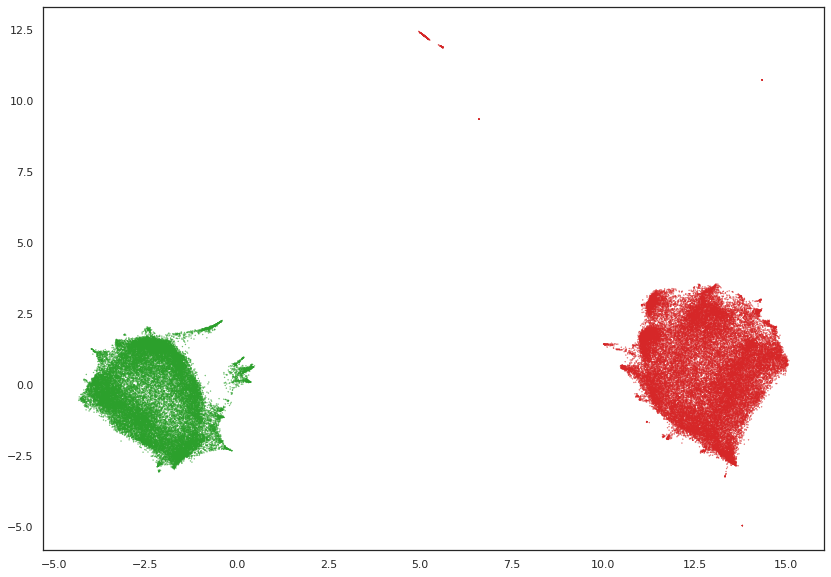}}

    \stackunder[5pt]{\includegraphics[width=0.6\linewidth]{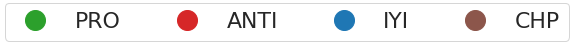}}{}
    
    \caption{Clusters of TD users for different topics. Labels of the manually annotated users are used to color-code the clusters. Colors not shown in the legend refer to clusters that have no manually annotated users.}
    \label{fig:targetClusters}
\end{figure*}

\begin{table}
    \centering
    \begin{tabular}{l|c|c}
        Topic & pro-Erdo\u{g}an & anti-Erdo\u{g}an \\ \hline
        Arab & 78.78 & 88.83 \\
        CHP & 83.65 & 86.32 \\
        Erdo\u{g}an & 92.74 & 91.95 \\
        HDP & 82.34 & 84.85 \\
        PKK & 78.27 & 79.76 \\
        Syrian & 83.35 & 85.8 \\
        Trump & 83.62 & 82.58 \\
        USA & 84.19 & 85.91 \\
    \end{tabular}
    \caption{Cluster labels overlap with label propagation}
    \label{tab:overlapWithLabelProp}
\end{table}

\paragraph{Can we infer the stance on a topic from a different topic?}
One of the symptoms and reinforcing causes of polarization is ``biased assimilation'', where individuals readily accept confirming evidence and are rather critical when provided with disconfirming evidence~\citep{Dandekar_2013}. We inspected if this phenomenon implies that there is correlation between clusters on different topics. {First, we inspected the clustering overlap with the pro- and anti-Erdo\u{g}an labels obtained earlier using label-propagation on the ED dataset as reported in \textbf{Table~\ref{tab:overlapWithLabelProp}}. As the results show, the positions towards different topics highly-correlate with users' positions during the election period. The greatest overlap was between the position towards Erdo\u{g}an in the ED and TD datasets. We also plotted the projected users for the 8 topics, retaining the users for which we have gold labels (from Table~\ref{tab:proAntiErdogan}), in \textbf{Figure~\ref{fig:targetClusters}}. Again, we can see that for certain topics, we can observe fine-grained separation between groups. In \textbf{Figure \ref{fig:nmi}},} we report adjusted mutual information (AMI) heatmaps across various topics, {adjusted for randomness} \citep{nmi}. {Mutual Information is a measure of the dependence between two clustering solution, and AMI accounts for higher mutual information scores when the number of clusters is larger.}\footnote{We used Scikit-Learn implementation of Adjusted Mutual Information \url{https://scikit-learn.org/stable/modules/generated/sklearn.metrics.adjusted_mutual_info_score.html#sklearn.metrics.adjusted_mutual_info_score}} 
We can see that topics influence similar stances towards other topics such that the minimum AMI score in the table is 0.70. AMI score is especially high for particular topic pairs, such as Europe and USA (0.97), Kurdish and USA (0.95),  USA and Trump (0.95), and Syrian and Arab (0.89). Overall, our approach enables investigation of correlation between stances for different issues.
\paragraph{What does polarization quantification inform us on different topics?}
We computed the polarization between the user clusters using Random Walk Controversy (RWC) measure \citep{darwish2019quantifying,garimella2018quantifying}. Given a graph of connected users, where the nodes denote users and the weights of the edges denote the user cosine similarities, computed based on retweets, RWC computes the shortest graph traversal from a set of random users to prominent users either with the same or different stances, {where prominent nodes are the top $n$ nodes with the most connections in a community \citep{darwish2019quantifying}. Two nodes are considered connected if their cosine similarity is not zero. The score is formulated as: $RWC = P_{AA} P_{BB} - P_{AB} P_{BA}$, where $A$ and $B$ are different classes and $P_{XY}$ is the probability that a random node in $X$ would reach a highly connected node in $Y$.}

\begin{figure*}[!htb]
\centering
\subfigure[Erdogan (Pro)]{\includegraphics[width=0.18\textwidth]{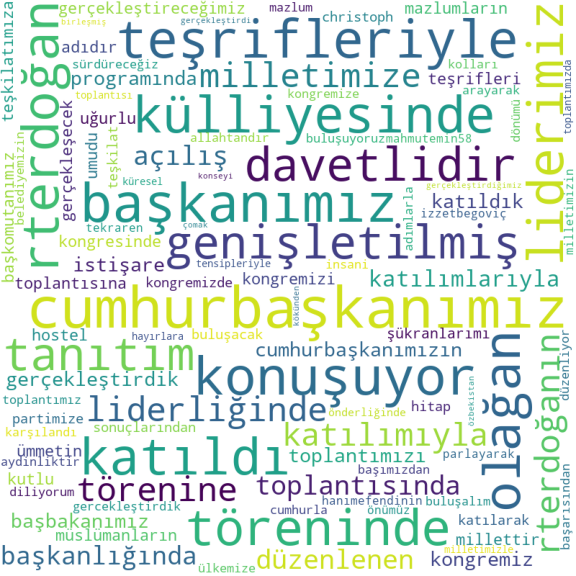}}\hfill
\subfigure[Erdogan (Anti)]{\includegraphics[width=0.18\textwidth]{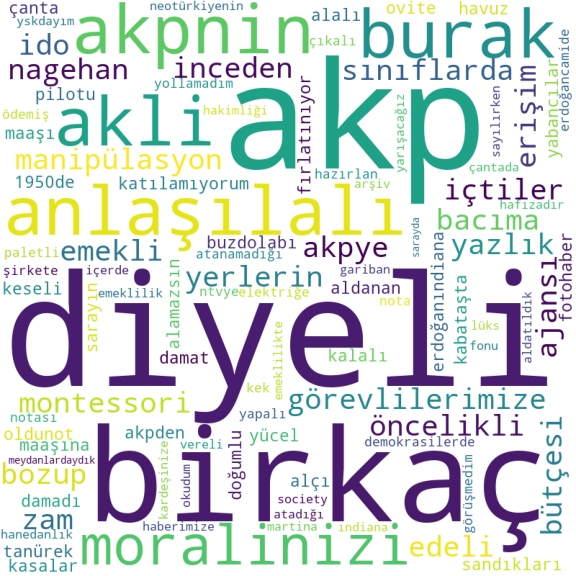}}\hfill
\subfigure[Arab (Pro)]{\includegraphics[width=0.18\textwidth]{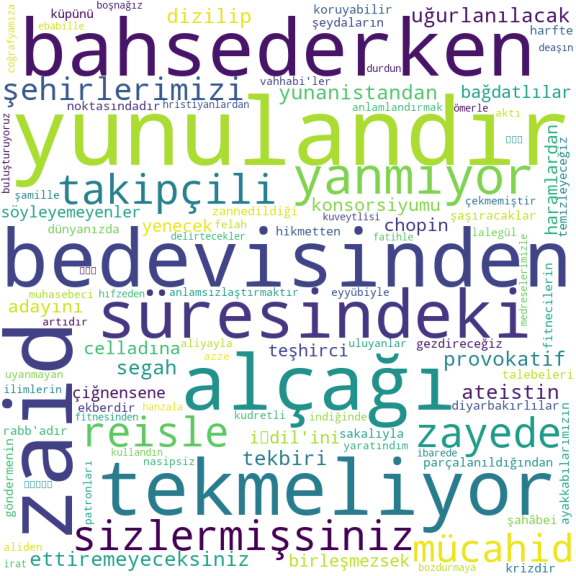}}\hfill
\subfigure[Arab (Anti)]{\includegraphics[width=0.18\textwidth]{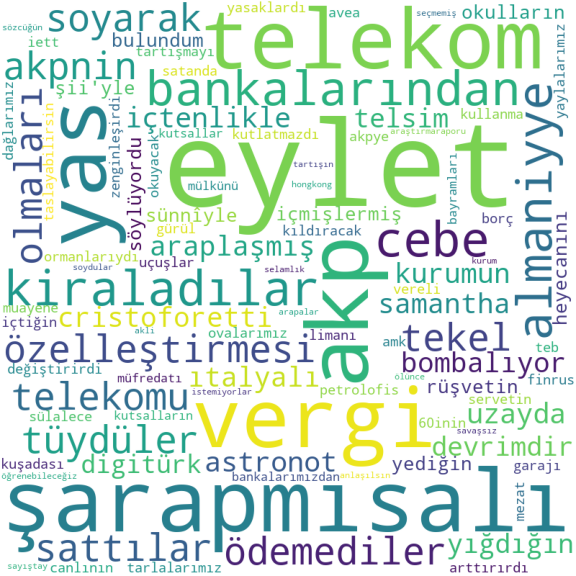}}
    \caption{Word Clouds Generated by prominent terms in each cluster for topics ``Erdo\u{g}an" and ``Arab".}
    \label{fig:WordClouds}
\end{figure*}
We computed RWC polarization measure on the aforementioned 8 target topics. \textbf{Table \ref{tab:RWCRes}} shows the RWC values, which range between 0 and 1, where 0 implies no polarization and 1 implies extreme polarization. {The results suggest that Turkish users are polarized on all topics. However, the stance of people towards particular issues, such as Erdo\u{g}an and HDP, cause more polarization than others. We also observe different RWC scores for similar topics. For instance, HDP causes more polarization than PKK, which is considered as a terrorist group by Turkey and USA. While HDP is a legitimate political party, many Turkish citizens want to ban HDP from participating in elections due to its alleged relation to PKK, yielding different stances towards HDP.}  
\begin{table}[tbp]
    \centering
    \begin{tabular}{l|c}
Target	&	RWC	\\ \hline
Arabs	&	0.81	\\
CHP	&	0.54	\\
Erdo\u{g}an & 0.89 \\
HDP	&	0.88	\\
PKK	&	0.71	\\
Syrians	&	0.69	\\
Trump	&	0.62	\\
USA	&	0.56	\\
    \end{tabular}
    \caption{RWC polarization measure across targets.}
    \label{tab:RWCRes}
\end{table}

\section{Qualitative Analysis and Discussion}\label{sec_discussion}
In this section, we conduct qualitative analysis to shed light on our approach's performance, and discuss its limitations. 

\subsection{Semantic Differences in Clusters}
For our analysis, we identify the most prominent terms in each cluster 
to show how people talk about the same issue in different contexts.
We assign \textit{prominence} scores for terms using valence scores~\citep{conover2011political} and term frequencies. 
The valence score of a term in a set of tweets $D_a$ captures the degree of its occurrence in $D_a$ compared to another set $D_b$.

We define the prominence 
score of a term in a set of tweets as the product of its valence score and its term frequency as follows: 

 \begin{equation}\label{eq:prominence}
     Pr(t, D_a, D_b) = \log{(tf_{t, D_a})}\times (2 \frac{\frac{tf_{t,D_a}}{|D_a|}}{\frac{tf_{t,D_a}}{|D_a|}+\frac{tf_{t,D_b}}{|D_b|}}-1) 
 \end{equation}
 \noindent
where $t$ is the term of interest, $D_a$ is a set of tweets (i.e., tweets of users in a cluster of interest), $D_b$ is the other tweet set, $tf_{t,D_a}$ and $tf_{t,D_b}$ denote the term frequency of $t$ in $D_a$ and $D_b$ respectively, and $|D_a|$ and $|D_b|$ are the sums of the frequencies of all the terms in $D_a$ and $D_b$ respectively.

\textbf{Figure~\ref{fig:WordClouds}} shows word clouds of the most prominent terms for the topic ``Erdo\u{g}an" and ``Arab". The word clouds of other topics are omitted due to space limitations.
We notice a remarkable contrast between how users in different clusters use distinctive terms towards a target. 

For instance, the top terms of ``Erdo\u{g}an'' topic from pro users include ``liderimiz" (our leader), ``ba\c{s}kan{\i}m{\i}z" (our leader), ``cumhurba\c{s}kan{\i}m{\i}z" (our president), ``te\c{s}rifleriyle" (with his honouring [by his visit]), and ``ba\c{s}komutan{\i}m{\i}z" (our commander-in-chief); whereas the top terms from the anti users include ``AKP''\footnote{non-official abbreviation of Erdo\u{g}an's party, mostly used by opposition not to call his party AKparti, which means White Party, suggesting innocence.}, words from popular phrases used for criticism (``diyeli", ``birka\c{c}"), words related to the 2013 Gezi protests (``i\c{c}tiler", ``bac{\i}ma", ``Kabata\c{s}'ta"),\footnote{A large protest against government in 2013.} and other issues such as the economy (``zam" (price increase), ``b\"{u}t\c{c}esi" (budget)) and allegations about Erdo\u{g}an's son (Burak).

Regarding the ``Arab" topic, top words of pro-Erdogan cluster include words about the political conflict between UAE and Turkey (e.g., ``Zaid" (UAE ruler) and ``Zayede" (to UAE ruler)), and complaints about discrimination against Arabs (``bahsederken" (expressing double standard)). However, the top words about Arabs in the anti-Erdo\u{g}an cluster include words about the sale of a state-owned telecom operator to a Lebanese businessman (e.g., ``Telekom", ``Telekomu", and ``\"{o}zelle\c{s}tirmesi" (privatization of public entities)).  We also observe similar contrast in other topics. For instance, regarding the topic ``Syria", top words in the pro-Erdo\u{g}an cluster include words about social integration of Syrian refugees, while top words in anti-Erdo\u{g}an cluster are about crimes committed by refugees. 

Overall, users in different clusters raise completely different issues and used different terms regarding the same topics with opposite stances.

\subsection{Misclustered Accounts}
\begin{table*}[!htp]

    \centering
\begin{tabular}{p{5cm}|p{11cm}}
    \textbf{Possible Reason}  &  \textbf{Translation of a Sample Tweet} \\ \hline
    Non-Turkish phrases  & ``biji serok Erdo\u{g}an'' slogans in the skies of Diyarbakir   \\ \hline
  Challenges in Turkish Sentences
   & ``The dishonoured anchorman of STV insulted Erdo\u{g}an'' \\ \hline
Mentioning other political entities& ``the love for Erdo\u{g}an is more powerful than  PKK FETO ISIS DHKPC Angela Merkel CHP HDP PYD YPG Netherlands USA'' \\ \hline
Sarcastic tweets &  ``Erdo\u{g}an, if you will not go, then at least love us, even if it is a lie'' \\ \hline
Semantic ambiguity &  a tweet has two possible meanings: ``Erdo\u{g}an is hypnotising the dog'' and ``the dog Erdo\u{g}an is hypnotising'' \\ \hline

 Target Ambiguity & ``We are informed about the referendum process by our dear Party Province Leader Erdo\u{g}an Altan. We thank him.\\ \hline
 Criticism towards Supported Party &  
 ``He made a person who is fan of Erdo\u{g}an as a candidate against Erdo\u{g}an. But he is still the leader of CHP with no shame" \\ \hline 
\end{tabular}
\caption{Possible reasons of misclustering with examples.}
    \label{tab:qual}
\end{table*}

Next, we investigated the misclustered users (e.g., pro-Erdo\u{g}an accounts clustered with anti-Erdo\u{g}an accounts). We randomly picked 15 pro-Erdo\u{g}an and 15 anti-Erdo\u{g}an accounts that were misclustered. We manually inspected their tweets (549 and 230 tweets respectively) about Erdo\u{g}an in order to understand the possible reasons for errors. While in most of the cases the political stances of the misclustered users were clear, there were many linguistic challenges and Turkish-specific political issues that might have potentially caused such misclustering. \textbf{Table~\ref{tab:qual}} shows sample tweets (demonstrating such challenges) translated into English.\footnote{Due to excessive slang, the translations are not necessarily literal.}

The main causes of errors were:

 \noindent
      \textbf{Non-Turkish Phrases:} We observed that Turks living abroad also tweet about Erdo\u{g}an in foreign languages. These non-Turkish words might adversely affect clustering. The sample tweet shown in \textbf{Table~\ref{tab:qual}} also shows a code-switching challenge where users tweet in Turkish but also use Kurdish phrases (``biji serok'' means ``Long Live''), which is actually mostly used by PKK supporters for their leaders.
     
     \noindent
      \textbf{
      Turkish Linguistic Complexity:} Semantic analysis of Turkish sentences is especially hard due to its rich and complex morphology. It is an agglutinative language yielding long words with many morphemes. As an extreme but popular example, the word "uygarla\c{s}t{\i}ramad{\i}klar{\i}m{\i}zdanm{\i}\c{s}s{\i}n{\i}zcas{\i}na" can be translated into English with a phrase of 12 words ``as if you are one of those whom we could not civilize". Moreover, the sentences have a free word-order sentence structure, where words can be in any order. For example, a headword of a noun phrase can come before or after the other words in the phrase. Perhaps due to such Turkish-specific linguistic challenges, our approach fell short in ``understanding'' some Turkish sentences. For instance,  we observed that 9 out of the 15 misclustered pro-Erdo\u{g}an accounts insulted people other than Erdo\u{g}an, and free word order potentially confused the classifier.   

       We have also observed that spam tweets, which use hashtags related to Erdo\u{g}an, cause misclustering.

       \noindent
       \textbf{Mentioning Other Political Entities:} We have observed that 
       some accounts mention names of political entities frequently, which might increase their topical similarity with those groups, eventually causing misclustering. For instance, a pro-Erdo\u{g}an user expressed that he is an Ataturkist. However, people identifying themselves as Ataturkist are more frequently CHP supporters than Erdo\u{g}an supporters, as Atat\"{u}rk founded the CHP party.  
       
       \noindent
       \textbf{Sarcastic Tweets:} Sarcasm is frequently used in political discussions, and we observed many sarcastic tweets in our analysis. For instance, anti-Erdo\u{g}an accounts may sarcastically refer to Erdo\u{g}an as ``the leader of Muslims'' and ``the hope of oppressed people''. 
       
       \noindent
       \textbf{Semantic Ambiguity:} Ambiguity is one of the main challenges in NLP applications. We observed a tweet from a pro-Erdo\u{g}an account which has two possible meanings (See \textbf{Table~\ref{tab:qual}}), with one of the possible meanings expressing extreme negative sentiment towards Erdo\u{g}an.
       
       \noindent
       \textbf{Target Ambiguity:} Erdo\u{g}an is a popular name and surname in Turkey. Since we do keyword matching to find tweets, some of the tweets obtained are actually about other people whose surnames are also Erdo\u{g}an.

       \noindent
        \textbf{Stance Ambiguity:} We observed that an anti-Erdo\u{g}an user had tweets about Erdo\u{g}an that do not express his stance towards Erdo\u{g}an directly, even though he explicitly states his support for CHP in his profile.
        
        \noindent
        \textbf{Hashtag Hijacking:} While a political group uses a specific hashtag to promote their political message, people from other political groups may use that particular hashtag to promote their own messages.  We also observed this behavior in our data where anti-Erdo\u{g}an accounts use pro-Erdo\u{g}an hashtags to tweet against Erdo\u{g}an.  
        
        \noindent
        \textbf{Criticism of Supported Party:} Supporters of a political party might also criticize specific people in their party, as shown in \textbf{Table~\ref{tab:qual}}. 
       
       \noindent
       \textbf{Exposing Negative News about the Target:} In 6 out of 15 anti-Erdo\u{g}an accounts, we observed that people share just negative news about Erdo\u{g}an without any personal comments. The lack of personal comments and background information complicates stance detection.
       
       \noindent
       \textbf{Political Alignments:} Though AKParti and MHP aligned together for the 2018 election, MHP and CHP aligned together in 2014 against Erdo\u{g}an. Thus, supporters of MHP may have had different stances at different time periods. 
       One such example of this is a pro-MHP user, who heavily criticized Erdo\u{g}an in older tweets, while suggesting that he voted for Erdo\u{g}an in later ones. 

\section{Conclusion}

In this work, we explored the polarization between Twitter users who either supported or opposed Erdo\u{g}an in the 2018 Turkish elections using a novel fine-grained content-based unsupervised stance detection method. We collected 108M tweets posted during the period leading up to the election, and we used a semi-supervised method to label 653K users, who posted 60M tweets, as pro- or anti-Erdo\u{g}an. Subsequently, we randomly selected 168K labeled users and crawled their timelines, collecting 213M tweets in total, covering the period before and after the election.
For our analysis, we employed a novel stance detection method, which uses subword-level  embeddings to represent users based on the content of their tweets on a particular topic. Using subword-level embeddings helped deal with morphological complexities of Turkish. Next, our method projected users onto a lower dimensional space, bringing similar users closer together and pushing dissimilar users further apart, and then clustered the users. We observed that our method can be used to detect fine-grained stances of users towards different topics with high accuracy. We applied our method to cluster user stances towards various polarizing issues in Turkish society, noting correlations between positions across topics. Additionally, we used the resultant clusters to quantify polarization on topics of interest.

\bibliography{references}
\end{document}